\documentclass[useAMS,usenatbib]{mn2e}

\usepackage[T1]{fontenc}

\usepackage{txfonts}
\usepackage{graphicx}
\usepackage{amssymb}

\newcommand{\eqn} [1] {
\begin{equation}
#1
\end{equation}}

\def\ds {$\delta$ Scuti}
\def\dss {$\delta$ Scuti stars}
\def\Hip {{\emph{Hipparcos~}}}
\def\mv {{M_{\mathrm{v}}}}

\def\bv {{B_2-V_1}}
\def\teff {{T}_{\mathrm{eff}}}

\def\vsini {{\vr\!\sin\!i}}
\def\vr {{V}}

\def\muHz {{\mu\mbox{Hz}}}
\def\kms {{\mathrm{km}\,\mathrm{s}^{-1}}}
\def\msol {{\mathrm{M}_\odot}}

\def\rsol {{\mathrm{R}_\odot}}
\def\BU {{BU\,Cnc}}
\def\BN {{BN\,Cnc}}
\def\BW {{BW\,Cnc}}
\def\BS {{BS\,Cnc}}
\def\BV {{BV\,Cnc}}

\def\anl {{\alpha_{\mathrm{NL}}}}
\def\amlt {{\alpha_{\mathrm{MLT}}}}

\def\michel {{M99}}
\def\ph  {{PH99}}

\def\ppa {{Physical Process in Astrophysics}}
\def\apj{ApJ}%
\def\aap{A\&A}%
\def\aaps{A\&AS}%
\def\mnras{MNRAS}%
\title[Mode stability in \dss: linear analysis versus
          observations in open clusters]{Mode stability in \dss: linear analysis versus
          observations in open clusters}
\author[J.C. Su\'arez, E. Michel, G. Houdek, F. P\'erez Hern\'andez, and Y. Lebreton]{J.C. Su\'arez$^{1,2}$\thanks{Associate researcher at institute (2), E-mail:jcsuarez@iaa.es}, E. Michel$^{2}$, G. Houdek$^{3}$, F. P\'erez Hern\'andez$^{4,5}$, Y. Lebreton$^{6}$\\
$^{1}$Instituto de Astrof\'{\i}sica de Andaluc\'{\i}a (CSIC), Granada, Spain \\    
$^{2}$LESIA, Observatoire de Paris-Meudon, UMR8109, Meudon, France \\              
$^{3}$Institute of Astronomy, University of Cambridge, Cambridge CB30HA, UK \\             
$^{4}$Instituto de Astrof\'{\i}sica de Canarias (IAC), Tenerife, Spain \\          
$^{5}$Departamento de Astrof\'{\i}sica, Universidad de La Laguna, Tenerife, Spain\\
$^{6}$GEPI, Observatoire de Paris-Meudon, Meudon, France}

\begin{document}

\date{Accepted . Received ; in original form }

\pagerange{\pageref{firstpage}--\pageref{lastpage}} \pubyear{2002}

\maketitle

\label{firstpage}

\begin{abstract}
             A comparison between linear stability analysis and observations of pulsation 
             modes in five \dss, belonging to the same cluster, is presented. The study is based on 
	     the work by \citet{MiHer99}, {\bf in which such a comparison was performed for a 
	     representative set of model solutions obtained independently for each 
	     individual star considered. 
	     In this paper we revisit the work by \citet{MiHer99} following, however, a
	     new approach which consists in the search 
	     for a single, complete, and coherent solution for all the selected stars,
	     in order to constrain and test the assumed physics describing these objects.
	     To do so, refined descriptions for the effects of rotation on the
	     determination of the global stellar parameters and on the adiabatic 
	     oscillation frequency computations are used.
	     In addition, a crude attempt is made to study the role of rotation on the
	     prediction of mode instabilities.
	     The present results are found to be comparable with those reported by 
	     \citet{MiHer99}. Within the temperature range $\log T_{\rm eff}$ = 3.87-3.88 
	     agreement between observations and model computations of unstable modes is restricted to 
	     values for the mixing-length parameter $\anl\simeq1.50$. This 
             indicates that for these stars a smaller value for $\anl$ 
             is required than suggested from a calibrated solar model.} 
	     We stress the point that the linear stability analysis used in this work
	     still assumes stellar models without rotation and that further developments 
	     are required for a proper description of the interaction between rotation 
	     and pulsation dynamics.
\end{abstract}


\section{Introduction}\label{sec:intro}

With spectral types from A2 to F0, the \dss\ are pulsating variables both in the main
sequence and in the sub-giant evolution phase (hydrogen shell burning). They are 
considered as good candidates for asteroseismic studies. Such stars show modes 
excited over a wide frequency range, including mixed modes known to be sensitive 
to the structure of the deep interior. These characteristics, together with the absence of 
magnetic field or metallicity peculiarities make them suitable for the asteroseismic 
study of hydrodynamical processes occurring in stellar interiors, such as convective 
overshooting, mixing of chemical species, and redistribution of angular momentum \citep{Zahn92}.
Recently, Su\'arez, Goupil \& Morel~(2006) reported that the effect of the distribution
of angular momentum on adiabatic oscillation frequencies in \dss\ is significant and that it can 
be detected by space missions such as \emph{CoRoT}. 
Due to the complexity of the oscillation spectra in \dss, their pulsating 
behaviour is not completely understood 
\citep[see][ for a complete review of unsolved problems in stellar pulsation physics]{Cox02}. 
In particular, great efforts are made nowadays to solve the problem of 
mode identification. Ground based multisite campaigns are regularly organised within 
coordinated networks, e.g.: STEPHI \citep{Michel00stephi} or DSN 
\citep{Breger00,Handler00}. Only a few tens of the low-degree modes 
have been detected from ground-based observations, with a maximum of around 75 modes for
the \ds\ star FG~Vir \citep{Breger05}. The recently launched space mission
\emph{CoRoT}\footnote{http://corot.oamp.fr} \citep{Baglin02} represents a unique 
opportunity for investigating such stars considering its much lower detection 
threshold and for obtaining data from quasi-uninterrupted time series 
over 5 months.

Additional uncertainties arise from the effect of rapid rotation, both directly on the 
hydrostatic balance in the star and through mixing caused by circulation or instabilities 
induced by rotation. 
The \dss\ are commonly fast rotators ($100<\vsini<200\,\kms$), and consequently the
symmetry of the multiplets is broken by the rotation. In 
the framework of a perturbation analysis, the second order effects induce
strong asymmetries in the splitting of multiplets \citep{Saio81,DG92} and frequency shifts 
must not be neglected even for radial modes \citep{Soufi95}.
\begin{table}
  \begin{center}
    \caption{Detected frequency peaks in Praesepe target stars.}
    \vspace{1em}
    \renewcommand{\arraystretch}{1.2}
    \begin{tabular}[h]{ccc|ccc}
      \hline\hline
        Star & &$\nu~(\muHz)$ & Star &  &$\nu~(\muHz)$ \\
      \hline
       \BU & $\nu_1$ & 193.3 & \BW & $\nu_1$ & 68.3 \\
           & $\nu_2$ & 195.1 &     & $\nu_2$ & 138.7\\
           & $\nu_3$ & 200.9 &     & $\nu_3$ & 139.1\\
           & $\nu_4$ & 215.5 &     & $\nu_4$ & 171.8\\
           & $\nu_5$ & 228.8 &     & $\nu_5$ & 223.0\\
           & $\nu_6$ & 229.9 &     & $\nu_6$ & 260.6\\
           &         &       &     & $\nu_7$ & 261.5\\
           &         &       &     & $\nu_8$ & 307.2\\
       \BN & $\nu_1$ & 263.6 &     & $\nu_9$ & 361.2\\
           & $\nu_2$ & 266.5 &     &         &      \\
           & $\nu_3$ & 279.5 & \BS & $\nu_1$ & 179.7\\
           & $\nu_4$ & 298.1 &     & $\nu_2$ & 197.2\\
           & $\nu_5$ & 300.2 &     & $\nu_3$ & 396.2\\
           & $\nu_6$ & 327.2 &     &         &      \\
           &         &       & \BV & $\nu_1$ & 190.4\\
           &         &       &     & $\nu_2$ & 193.6\\
           &         &       &     & $\nu_3$ & 380.9\\
           &         &       &     & $\nu_4$ & 234.3\\
      \hline
      \end{tabular}
    \label{tab:freqPrae}
  \end{center}
\end{table}
In this context, \citet{MiHer99}, hereafter \michel, proposed a technique to estimate 
and correct the effect of fast rotation on the determination of fundamental parameters 
for stars in clusters. Applying this technique to a set of \dss\ belonging to the 
Praesepe cluster, the authors showed that it was possible to reach a reasonable agreement 
between ranges of observed oscillation modes and overstable radial modes predicted by a 
linear stability analysis. 
However, agreement was found only for certain values of $\anl$, the mixing-length 
parameter that was used in the non-local time-dependent convection treatment of the 
stability computations. This implies that models for convective heat transport in the outer 
stellar layers may be calibrated against observations.

In \michel, different series of stellar model solutions {\bf were obtained, which assumed a 
wide range of physical parameters, e.g. overshooting, mixing-length parameter, metallicity, 
stellar age, etc. Considering these results as our reference domain of possible solutions,
the present work revisits the work by \michel\ but adopts an improved approach}. The main
idea is to search for one particular solution that explains the whole set of observations,
instead of sets of individual solutions for each star. To do so, refined techniques 
for modelling intermediate mass stars are {\bf required}, in particular such techniques that take
different effects of rotation into account. {\bf In concrete terms, in \michel\ the photometric 
parameters are corrected for the effect of rotation using the results of \citet{MaederPey70} 
who applied the von Zeipel gravity-darkening law to estimate the emergent flux of 
a rotating star. In the present work we follow the method described by \citet{Pe99} 
(hereafter \ph) which improves such a calculation by means of, among other modelling aspects, 
updated models and consider the gravity-darkening law given by \citet{Claret98},
which, unlike the von Zeipel law, is also valid for non-fully radiative stars.
Moreover, in \michel\ non-rotating equilibrium models were used	whereas here we employ 
pseudo-rotating models. Such pseudo-models take first-order effects of rotation into account by means 
of an effective gravity (see Sect.~\ref{ssec:stelparam}). One of the major improvements with 
respect to \michel's work lies in the computation of adiabatic oscillation frequencies. 
In \michel\ the effects of rotation
on the oscillation frequencies (up to the second order) are included in the manner
of \citet{PeHe95}, considering the perturbed frequency $\nu_n^\prime$ as
\eqn{\nu_{n}'=\nu_{n}+(A_{n}+B)\frac{\nu_{\mathrm{rot}}^2}{\nu_{n}}~,}
where $\nu_{n}$ is the unperturbed frequency, $\nu_{\mathrm{rot}}$ is the rotation
frequency, $B$ is a constant with an asymptotic value of $5/3$, and
$A_{n}$ corresponds to a coefficient depending on the eigenfunctions, and thus,
on the stellar model. These coefficients are obtained by interpolating in the
values computed by \citet{Saio81} for an $n=3$ polytrope. 
In the present work, following \citet{DG92} and \citet{Soufi98} 
(see also Su\'arez, Goupil \& Morel~2006) a complete treatment of 
second order effects (including near degeneracy) is used. Furthermore, this formalism takes 
the effect of the star deformation due to rotation into account.

With these improvements a similar methodology than that adopted in \michel\ is 
used here to study five \dss\ of the Praesepe cluster, four of which have already
been included in the sample considered by \michel. 
The fundamental parameters of these stars are determined from taking into account 
the effect of rotation on the photometric
observables using the updated method by \ph. These fundamental parameters
are then used to build representative asteroseismic models for each star, consisting of
pseudo-rotating equilibrium models (instead of non-rotating equilibrium
models used in \michel) and their corresponding adiabatic oscillation spectra
(properly corrected for the effect of rotation). These asteroseismic models
are then used to determine the \emph{observed} ranges radial orders, which,
as it was done in \michel, are then confronted with mode instability predictions obtained
from a linear stability analysis. We note that, as in \michel, the instability predictions
are carried out using equivalent envelope models which do, however, not take the effect of
rotation into account. This is so because, up to date, there are no reliable theories 
available which describe the effect of rotation on mode stability. Nevertheless, 
a crude estimate of this effect is addressed, which assumes that mode stability
depends predominantly on the effective temperature of the model \citep{Alosha75}.}

The paper is structured as follows: In Section~\ref{sec:obs} the main characteristics of the 
Praesepe cluster and the observational material are presented. Equilibrium models and 
isochrone computations are discussed in Section~\ref{sec:modelling}. Different aspects of the 
oscillation computations are described in Section~\ref{sec:oscilcomp}. In Section~\ref{sec:Compar}, 
the consistency of our solutions with the results obtained
by \michel\ is examined. Then, a detailed discussion of the present results
is given.
Finally, conclusions and perspectives are summarised in 
Section~\ref{sec:conclu}.

\section{Observational data}\label{sec:obs}

The selection of stars considered here contains five \dss\ belonging to the
Praesepe cluster. Four of these stars: \BW, \BS, \BU\ and \BN\ (HD\,73798, 
HD\,73450, HD\,73576, and HD\,73763 respectively) were 
observed by several campaigns of the STEPHI network \citep{Michel95}. The fifth star, 
\BV\ (HD\,73746), was observed by \citet{Frandsen01}.
The observed frequencies of these stars are listed in Table~\ref{tab:freqPrae}.

Regarding the cluster parameters, its distance modulus ($6.28\pm0.12$~mag)
was taken from \citet{Rob99} who derived it from mean cluster parallaxes computed from 
\Hip\ intermediate data. We assume a metallicity value $[M/H]$ of 0.170  which was 
derived from Geneva photometry of single stars with spectral types in the range F4-K3 
\citep{Gre00,Rob99}.
\begin{table}
  \begin{center}
    \caption{Photometric data for the five selected \dss. The columns represent
             name, HD number, absolute magnitude (in mag), $\bv$ colour index (in mag), 
	     and the projected velocity $\vsini$ (in $\kms$) of the star.}
    \vspace{1em}
    \renewcommand{\arraystretch}{1.2}
    \begin{tabular}[h]{cccccc}
      \hline\hline
        Star &  HD& $\mv$  & $\bv$ &  $\vsini$                \\
      \hline
      \BW\   & HD\,73576 & 2.18 & 0.063 & 170 & \\
      \BS\   & HD\,74763 & 2.19 & 0.048 & 135 & \\
      \BV\   & HD\,73798 & 2.36 & 0.079 & 110 & \\
      \BU\   & HD\,73450 & 1.37 &-0.004 & 205 & \\
      \BN\   & HD\,73746 & 1.52 & 0.016 & 200 & \\
      \hline
      \end{tabular}
    \label{tab:photclusters}
  \end{center}
\end{table}
Apparent magnitudes and indices were taken from the Rufener's catalogue 
\citep{Ruf88}. In Table~\ref{tab:photclusters} the specific photometric data
for the five selected \dss\ are listed. The projected velocity values ($\vsini$)
were taken from \citet{Royer02}.

\section{Modelling}\label{sec:modelling}

%
\begin{figure}
  \includegraphics[width=9.2cm]{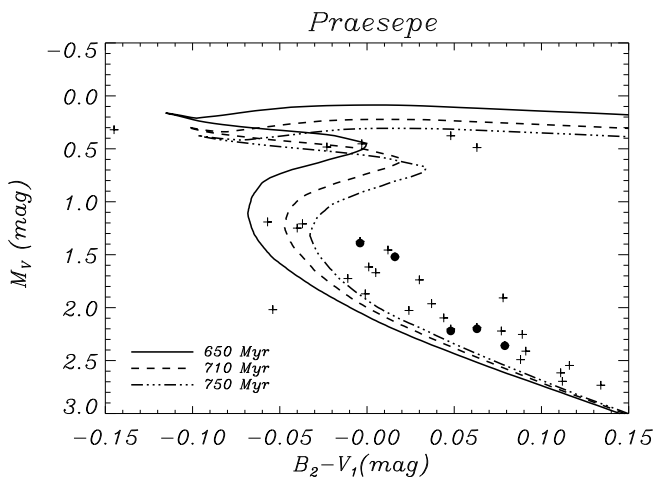}
  \caption{Colour -- magnitude diagram for the Praesepe cluster. Crosses
           represent the observations of stars with confirmed membership. Filled
	   circles correspond to the \dss\ selected for the present work. 
	   All objects are corrected for distance. Curves represent
	   isochrones of relevant ages computed from models with
	   $\amlt=1.614$ and $d_\mathrm{ov}=0.2$.}
  \label{fig:isocStd650-750Myr}
\end{figure}
In the process of searching representative models for the five selected \dss, 
we take advantage of their cluster membership in several aspects: firstly, the chemical 
composition and age are assumed to be the same (to a relatively good extent) for all 
objects. In addition, metallicities and distances are commonly estimated better
for stars in clusters than for field stars. Secondly, the photometric parameters
of target stars are corrected for the effect of rotation following \michel\ and 
\ph. Such corrections also provide improved estimates for the stellar parameters 
required for the modelling.

\subsection{Stellar models and isochrones}\label{ssec:isoc}

To characterise theoretically the observed stars within the cluster we first construct standard 
non-rotating models and isochrones 
that take into account the observed parameters presented in Section~\ref{sec:obs}, and
second, we also build models including uniform rotation (pseudo-rotating models) as described
in Section~\ref{ssec:stelparam}. The evolutionary 
stellar models are computed with the CESAM code \citep{Morel97}.   
Opacity tables are taken from the OPAL package \citep{Igle96}, complemented at 
low temperatures ($T\leq10^4\,K$) by the tables provided by \citep{AlexFergu94}. The 
atmosphere is constructed from a grey Eddington~$T-\tau$ relation. 
The Praesepe's metallicity value of $Z/X$ is derived from the $[M/H]$ value given 
in Section~\ref{sec:obs}, assuming $(Z/X)_\odot=0.0245$ \citep{GrevesseNoels93}, 
$Y_{\mathrm{pr}}=0.235$, and $Z_{\mathrm{pr}}=0$ for the helium and heavy element 
primordial mass abundances, and the value $\Delta Y/\Delta Z=2$ for the enrichment ratio. 
This yields an initial helium abundance $Y=0.285$ and a heavy element abundance $Z=0.025$ 
for the cluster. Convection is treated with a local mixing-length model \citep{BohmVit58} 
and we assume various values for the mixing-length parameter $\alpha_{\rm MLT}=l/H_{\rm p}$, 
where $l$ is the mixing length and $H_{\rm p}$ is the pressure scale height. {\bf In particular,
values between 0.5 and 2, including the calibrated solar value of $\alpha_{\rm MLT}=1.614$, 
are considered}.  For the overshoot parameter 
$d_{\rm OV}=l_{\rm OV}/H_{\rm p}$ ($l_{\rm OV}$ being the penetration length of the convective 
elements) we assume values between 0.0 and 0.3.

For the construction of non-rotating isochrones, various sets of evolutionary sequences were computed 
for masses between $1\,\msol$ and $5\,\msol$, evolved from the zero-age main sequence (ZAMS) to the 
sub-giant branch. Isochrones were then obtained using the Geneva Isochrone Code.
Three representative isochrones around the age of the Praesepe cluster 
are depicted in Fig.~\ref{fig:isocStd650-750Myr}, after having previously transformed
their effective temperatures and luminosities into the Geneva photometric system. 
This transformation is performed using the calibration proposed by \citet{Schm82} for $\mv$, 
and by \citet{Kunzli97} for $(\bv)$. Potential binarity and the
effect of fast rotation, discussed in Section~\ref{ssec:stelparam}, are expected to induce systematic 
shifts of the models towards higher luminosities and lower effective temperatures when compared with single 
non-rotating stars. Therefore, in order to avoid both effects, the fit of isochrones has to be made 
by adjusting the isochrones to the bottom envelope of the cluster in the colour--magnitude 
diagrams (hereafter CM diagrams).

Considering the physical parametrisation described above, a range of ages between 650 Myr and 
750 Myr is then found to be representative of the cluster (Fig.~\ref{fig:isocStd650-750Myr}). 
This range is in agreement with values found earlier in the literature for the Praesepe cluster.
The age will be adjusted more carefully in Section~\ref{ssec:stelparam}.

\begin{figure}
  \includegraphics[width=9.2cm]{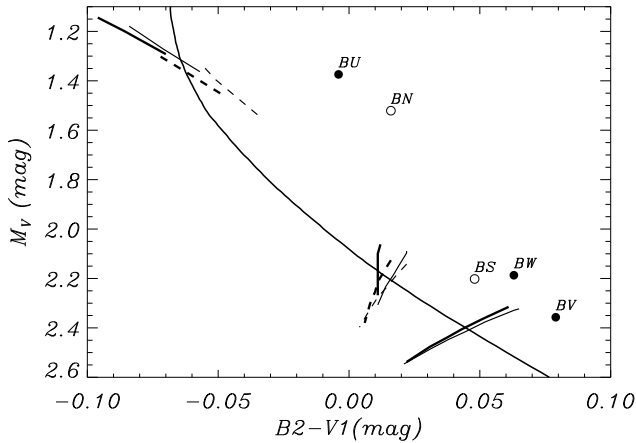}
  \caption{Colour -- magnitude diagram showing the corrections (two segments per star) of the 
           observed photometric parameters for the effect of rotation.
           Circles represent the observed \dss. {\bf Dashed and continuous segments correspond to 
	   empty and filled circles, respectively. The three pairs of segments
	   on bottom are associated with \BS, \BW, and \BV. The two remaining
	   ones on top are associated with \BU\ and \BN. Thick segments correspond
	   to corrections obtained from an increase of $\vsini$ by +10\%, and the thin segments
	   correspond to corrections obtained from decreasing $\vsini$ by -10\%.}
	   The continuous curve corresponds to a
	   650 Myr isochrone (see Fig.~\ref{fig:isocStd650-750Myr}).} 
  \label{fig:corr}
\end{figure}
%
\subsection{Stellar parameters corrected for rotation effects}\label{ssec:stelparam}

The modelling of the observed \dss\ requires the knowledge of
certain basic stellar parameters including the values for mass, effective 
temperature and luminosity (and/or gravity).
Two of these quantities can be determined from high resolution 
spectroscopy or, as in the present case, by means of photometric
observations. However, as shown in \michel, fast rotation must be
taken into account when estimating fundamental parameters from
photometric indices. In order to take this into account, the authors 
also proposed a method for 
correcting the photometric indices for the effect of fast rotation.
Here we use an improved version of this method, developed by \ph\
{\bf who considered three main effects: Firstly, the shape distortion 
caused by rotation is such that a star seen pole-on has a larger 
projected area and hence looks brighter than if it were seen equator-on. 
Secondly, the flux (and hence the effective temperature) is larger 
at the poles than at the equator (gravity-darkening effect), and 
thirdly, rotation decreases the intrinsic luminosity and increases 
the mean radius of the star. As a result of these effects the location
of a rotating star in a colour-magnitude diagram (and thereby in an
HR diagram) depends not only on the angular velocity 
of the star but also on its angle of inclination $i$.  
The basic idea of the method is to obtain, for a given rotating star
and for a given $\vsini$ value, a segment covering the 
different potential positions in the colour-magnitude diagram (i.e. for the 
corresponding putative stars which are not rotating). The \emph{corrections} 
so obtained for the selected \dss\ are displayed in Fig.~\ref{fig:corr}, in
which the positions of the non-rotating co-partners are varied with the angle 
of inclination $i$, and with the rotation rate ${\omega}$, defined as 
${\omega}=\Omega/\Omega_{\mathrm{c}}$, where $\Omega$ is the angular 
rotational velocity of the star, and $\Omega_{\mathrm{c}}$ is its break-up 
rotational velocity.\footnote{This corresponds to the rotational velocity that the 
star would have if the centrifugal force is balanced by the gravitational attraction at 
the equator.}. Although the \emph{corrections} are model-dependent,
for a given isochrone, this dependence is weak and the uncertainty introduced
smaller than that coming from the observational error in $\vsini$. } 
An error on the $\vsini$ values of $\pm10$\% is assumed, which yields two segments per star. 

\begin{table*}
  \caption{Characteristics of the non-rotating counterparts for the selected \dss\
           of the Praesepe cluster. These characteristics are obtained
	   with the method for correcting the effect of rotation on photometric
	   parameters (Section~\ref{ssec:stelparam}). The indicated ranges
	   correspond to solutions, from left to right, for which 
	   $\vsini$ was varied by +10\% and -10\% from its observed value
	   (see more details in Section~\ref{ssec:stelparam}). 
	   The best results with this method are obtained for an age of 650~Myr for the cluster.
	   The different columns are: the star identification from the General
	   Catalogue of Variable Stars (GCVS), the absolute magnitude $\mv$; 
	   the colour index $\bv$; the mass $M$ (in units of $\msol$); 
	   the radius $R$ (in units of $\rsol$); 
           the effective temperature $\teff$ (log, in K); the gravity $g$
	   (log, in $\mbox{cm\,s}^2$); the rotation rate $\omega$ (in units of
	   the break-up frequency $\Omega_{\mathrm{c}}$); the cyclic rotation frequency $\nu_{\mathrm{rot}}$ (in $\muHz$), 
	   and finally the inclination angle $i$ of the star.}\vspace{1em}
    \renewcommand{\arraystretch}{1.2}
    \begin{tabular}[h]{ccccccccccc}
      \hline\hline
       ID & $\mv$  & $\bv$ & $M$ & $R$ & $\teff$
            & $g$ & $\omega$ & $\nu_{\mathrm{rot}}$ & $i$ \\
      \hline

      BU\,Cnc & 1.29--1.36  & -0.070\,--\,(-0.057) & 2.09\,--\,2.11 & 2.52\,--\,2.56  & 3.91\,--\,3.91
                            & 3.95\,--\,3.94  &  0.95\,--\,0.95 & 19.05\,--\,14.70  &  56.27\,--\,51.17\\
      BN\,Cnc & 1.36\,--\,1.35  & -0.062\,--\,(-0.055) & 2.05\,--\,2.05 & 2.42\,--\,2.42  & 3.91\,--\,3.91
                            & 3.98\,--\,3.98  &  0.90\,--\,0.80 & 18.93\,--\,16.83  &  90\,--\,66.30\\
      BW\,Cnc & 2.19\,--\,2.23  &  0.011\,--\,0.014    & 1.73\,--\,1.70 & 1.87\,--\,1.84  & 3.88\,--\,3.88
                            & 4.13\,--\,4.14  &  0.90\,--\,0.90 & 24.14\,--\,24.71  &  42.63\,--\,48.25\\
      BS\,Cnc & 2.21\,--\,2.24  &  0.011\,--\,0.014    & 1.73\,--\,1.70 & 1.86\,--\,1.83  & 3.88\,--\,3.88
                            & 4.13\,--\,4.14  &  0.80\,--\,0.80 & 20.04\,--\,20.35  &  45.47\,--\,40.22\\
      BV\,Cnc & 2.41\,--\,2.42  &  0.043\,--\,0.044    & 1.65\,--\,1.64 & 1.75\,--\,1.74  & 3.87\,--\,3.87
                            & 4.16\,--\,4.17  &  0.80\,--\,0.80 & 24.58\,--\,24.72  &  34.70\,--\,31.42\\
      \hline
    \end{tabular}
    \label{tab:corrStd}
\end{table*}
After an iterative process (more details in \ph) it is possible,
in most of the cases, to converge to an intersection between the isochrone
and the segments. The non-rotating co-partners of our observed stars closest 
to the intersection between the isochrone and the segments give us an estimate 
of both $i$ and $\omega$. Additionally, the isochrone models at the intersection 
provide estimates of the required remaining stellar parameters. 

As mentioned in \ph, the photometric correction for rotation depends on
the selected isochrone. In the present case,
the convection parameters $\amlt$ and $d_\mathrm{ov}$ were adjusted within the ranges given in
the previous section; the best solution (given by this correction) was obtained for
$\amlt=1.614$ and $d_\mathrm{ov}=0.2$, leading to an optimal age of 650~Myr for
the cluster.
This solution was selected under two main criteria: first, a standard fitting of
isochrones, and second, the existence of suitable photometric corrections
(see \ph\ and Su\'arez et al. 2002 for more details) for the whole sample of 
stars. Both criteria make the \emph{solution} converge rapidly to
an age of 650~Myr ($\pm20-40$ Myr). The age uncertainty of 20--40 Myr can be neglected
in terms of global characteristics of the non-rotating co-partners
\citep[see the influence of age on photometric corrections for rotation
in different open clusters in][]{Sua02aa}. 
For this solution, the following range of masses are
found: from $M=2.05$\,--\,$2.11\msol$ (our \emph{high mass} stars) to 
$1.64$\,--\,$1.65\msol$ (our \emph{low mass} stars), which is within the range of masses 
obtained in \michel. In particular, the \emph{high mass} models are found to be 
representative of \BU\ \& \BN; the $M=1.70$\,--\,$1.73\,\msol$ models are
representative of \BW\ \& \BS, and finally, the $M=1.64$\,--\,$1.65\,\msol$ models are
representative of \BV. The characteristics of the different models are 
summarised in Table~\ref{tab:corrStd}. We recall that the different ranges 
listed in Table~\ref{tab:corrStd} correspond to solutions in which $\vsini$ was
varied by $\pm10\%$ from its observed value. As explained in \ph\ and \citet{Sua02aa},
this uncertainty largely dominates the errors of the method.

We then compute stellar models including uniform rotation for the five considered stars,
adopting the corrected values for their masses and ages obtained from the method described above. 
To take first order effects of rotation into account, the equilibrium equations are modified in 
the CESAM code in the manner described in \citet{KipWeig90}. 
In particular, the spherically averaged contribution of the centrifugal 
acceleration is included by means of an effective gravity $g_{\mathrm{eff}}=g-{\cal A}_{c}(r)$,
where $g$ is the local gravity, $r$ is the radius, 
and ${\cal A}_{c}(r)=\frac{2}{3}\,r\,\Omega^2(r)$ is the centrifugal acceleration of matter 
elements. This spherically averaged component
of the centrifugal acceleration does not change the order of the hydrostatic equilibrium equations.
From now on we shall refer to such models as 'pseudo-rotating' models.
Although the non-spherical components of the centrifugal acceleration are not considered in
the equilibrium models, they are included in the {\bf adiabatic} oscillation computations by means of a linear 
perturbation analysis according to Soufi, Goupil \& Dziembowski~(1998)  
(see also Su\'arez, Goupil \& Morel~2006).
\begin{figure*}
  \includegraphics[width=8cm]{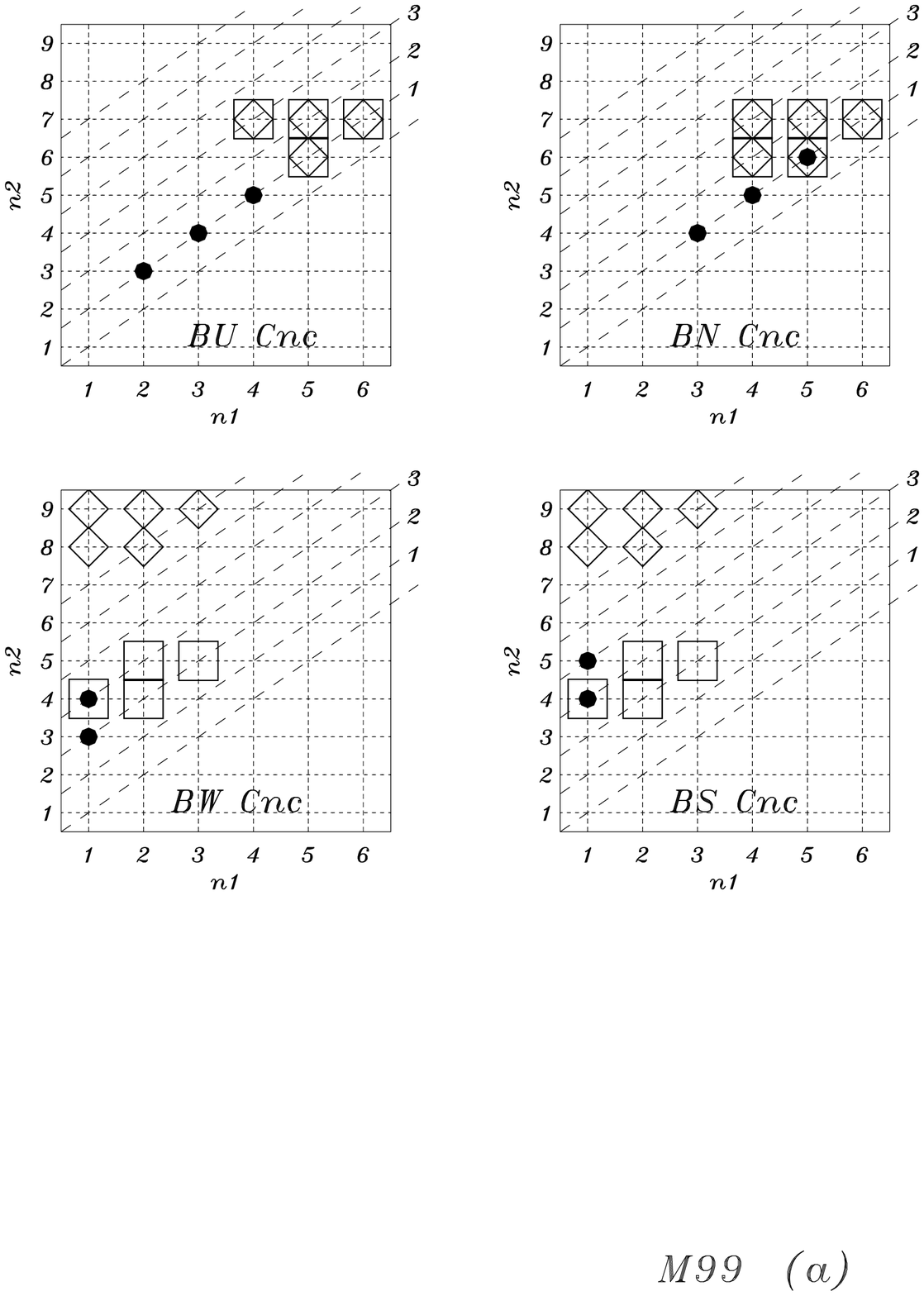}\hspace{1.50cm}
  \includegraphics[width=8cm]{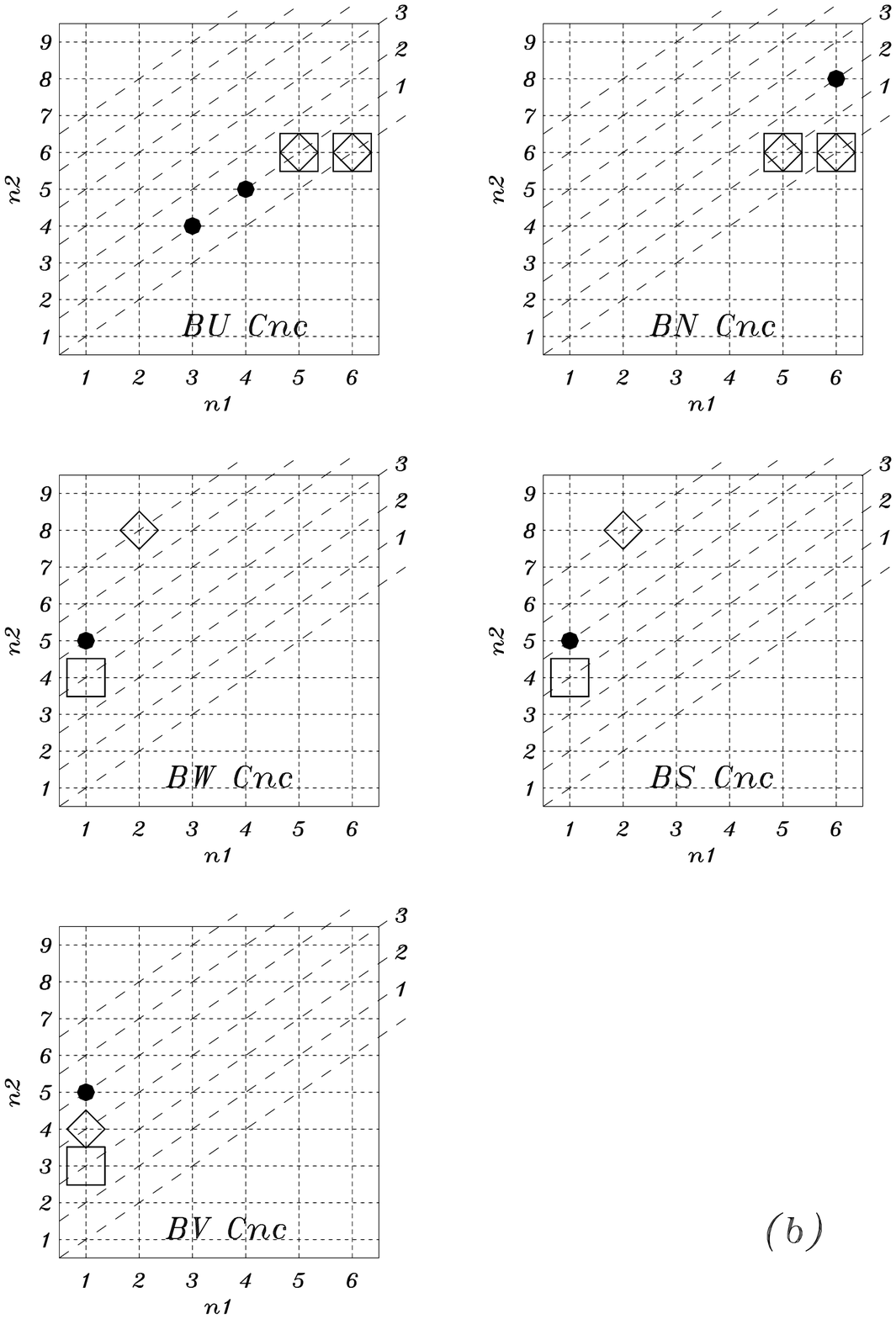}
  \caption{Observed and predicted (using linear stability analysis) ranges of unstable radial 
           modes for the selected \dss\ ($n_1$ is the lowest value, and
           $n_2$ the largest value of the radial order of the unstable modes). In panel~(a) 
           we re-plot the results obtained by \michel\ for \BU, \BN, \BS\
	   and \BW, and in panel~(b) the current results are displayed. 	   
	   Filled circles represent the observed ranges. Rhombus and squares
	   correspond to predicted radial order ranges for $\anl=1.89$ and
	   $\anl=1.50$, respectively. Each diagonal-dashed line represents the width (in
	   radial orders) of the represented ranges.}
  \label{fig:GammesPrae}

\end{figure*}
%

\section{Determination of observed and theoretical ranges of unstable modes}
\label{sec:oscilcomp}

We compute synthetic {\bf adiabatic} oscillation spectra corresponding to the pseudo-rotating evolutionary 
models described in Section~\ref{ssec:stelparam}. 
We restrict our computations to modes with degree $\ell\leq 2$ since
modes with $\ell\geq3$ are generally considered to be invisible in 
photometry (only visible in spectroscopy) for distant 
stars due to geometric cancellation. 
The ranges of observed radial and nonradial 
modes are similar to the predicted ranges of computed overstable radial modes.
This is to be expected, because driving and damping in \dss\
takes place predominantly in the HeII ionisation zone, which is rather close
to the stellar surface, where the vertical scale is much less than the
horizontal scale of the oscillations and when $\ell$ is low the modal inertia
is quite insensitive to degree $\ell$. It is therefore plausible to 
use only radial modes in the computations the results of which are also applicable to
modes of low degrees \citep[eg. ][]{Dziembowski01,Pagoda05}.
Then, proceeding as in \michel, for each star, the range of predicted 
unstable radial modes (according to linear growth rates) is compared with observed mode 
ranges.

\subsection{Determination of observed ranges of radial modes}\label{ssec:radord}

Theoretical {\bf adiabatic} oscillation spectra are computed with the oscillation code
{\sc filou} \citep{filou,SuaThesis}. 
As mentioned in Section~\ref{sec:intro}, we are using an improved version 
of \michel's treatment for second-order 
effects of rotation on the {\bf adiabatic} oscillation frequencies. 
{\bf In particular, we use the complete treatment of second-order effects of rotation
by \citet{Sua06rotcel}, which is based on the formalisms by \citet{DG92} and \citet{Soufi98}.
Furthermore, the theoretical adiabatic oscillation frequencies
are computed from the pseudo-rotating equilibrium models (Section~\ref{ssec:stelparam}), 
whereas \michel\ considered only non-rotating equilibrium models. We recall that these 
pseudo-rotating equilibrium models, representative of each star,
are computed from the rotationally corrected stellar parameters of 
Sect.~\ref{ssec:stelparam}.}

For each star the range of radial orders of the observed modes
is then determined in the manner of \michel, i.e. the locations of 
the peaks in the observed power spectrum are compared with the
locations of radial modes in the theoretical eigenspectrum of
the corresponding pseudo-rotating models. 

\subsection{Predicted unstable modes}\label{ssec:unstmodes}

The stability computations are carried out in the manner of \citet{Balmforth92}. 
We include the Lagrangian perturbations of the turbulent fluxes, i.e. the 
convective heat flux $\delta F_{\rm c}$ and turbulent pressure\footnote{
$p_{\rm t}=\rho\,w^2$ is the $(r,r)$-component of the Reynolds 
stress, where $\rho$ is the density of the mean stratification and $w$ is 
the rms vertical component of the turbulent velocity field.} 
$\delta p_{\rm t}$ in the stability computations 
according to the non-local, time-dependent mixing-length model by 
\cite{Gough77,Gough77proc}.
The equilibrium envelope models are obtained by specifying mass, luminosity and 
effective temperature as provided by the evolutionary computations of Section~3,
{\bf but they are not corrected for rotational effects}. 
The radiation field is treated in the Eddington approximation, and 
the atmosphere is considered to be grey and plane-parallel.
We adopt the nomenclature of M99, i.e. $\alpha_{\rm NL}=l/H_{\rm p}$ 
is the mixing-length parameter of the non-local convection model used in the 
stability computations. The non-local mixing-length parameter $\alpha_{\rm NL}$
is calibrated to the same depth of the outer convection zone as suggested 
by the evolutionary computations (see Section~\ref{sec:modelling}) 
which use the standard mixing-length 
formulation by \citet{BohmVit58} and a local mixing-length parameter 
$\alpha_{\rm MLT}=1.614$. The so obtained calibrated value of 
$\alpha_{\rm NL}=1.89$.
To study the effect of varying the mixing length on mode stability we
computed a second series of stellar models with $\alpha_{\rm NL}=1.50$.
Further details on the stability computations can be found in 
\citet{Houdek99}.

\section{Comparison between observed and predicted ranges of unstable 
         radial modes}\label{sec:Compar}

We proceed as in \michel\ and compare ranges of observed and predicted
radial orders $n$ of unstable modes for two values of $\anl$:
1.89 and 1.50 (see Section~\ref{ssec:unstmodes}). We recall that
this comparison does not imply that all observed modes are identified
as radial modes, but their frequency ranges are analysed in terms of unstable modes 
described within a range of radial modes (for details see \michel). 

The errors in the 
calibration from colour indices to effective temperatures can reach
$150\,\mbox{K}$. It is therefore adequate to accept an uncertainty of $\pm1\,n$ 
in the determination of the range of radial orders. In addition, this error also accounts for 
(1) the difference 
of ranges estimated in the co-rotating and the observer's frame, i.e., the rotational splitting;
for modes up to $\ell=2$, this represents a frequency range shift up to 
$\nu_{\mathrm{rot}}\,{\ell}\sim40\,\muHz$, with $\nu_{\mathrm{rot}}$ being the stellar rotational 
frequency (up to $\sim$ 20~$\mu$Hz); {\bf and (2), for the possible $\ell=1$ and 2 modes
with frequencies close to the boundaries of the radial order ranges.}

\subsection{Consistency with the \michel\ results}\label{ssec:Check}

As a first step, we examine the consistency of the present results with the results by
\michel. We do so by comparing the ranges of radial orders (observed and predicted ranges)
obtained here (Fig.~\ref{fig:GammesPrae}b) with the ranges reported by 
\michel\ (Fig.~\ref{fig:GammesPrae}a) for \BU, \BN, \BW\ and \BS\,. 
In both panels, the observed ranges of radial orders are represented by filled circles; those obtained
from the stability analysis using $\anl=1.89$ and $1.50$ are illustrated by rhombus 
and squares, respectively.
The coordinates ($n_1,n_2$) correspond to ranges of radial orders between $n_1$ and $n_2$. 
 
When performing this comparison, 
it has to be kept in mind that the present work 
includes various improvements in the stellar modelling over \michel\ and
consequently represents a more specific and more precise solution,
whereas the \michel\ results represent an extended and conservative 
set of solutions in terms of model parameters.
In this view, we are interested to check if our new solutions do belong
to the set of solutions by \michel\ (and if so, to what extent), 
or if they differ significantly.

Considering the observed ranges, Fig.~\ref{fig:GammesPrae} shows that the new solutions
lie within those reported by \michel. 
The only significant difference is observed for \BN\ and which has to be attributed to the
updated (high) $\vsini$ value adopted in the present work. 

As for the instability predictions, our results 
for the two considered values of $\anl$
are in general similar, within $\pm1\,n$, to the results reported by \michel.
\begin{figure}
  \includegraphics[width=9cm]{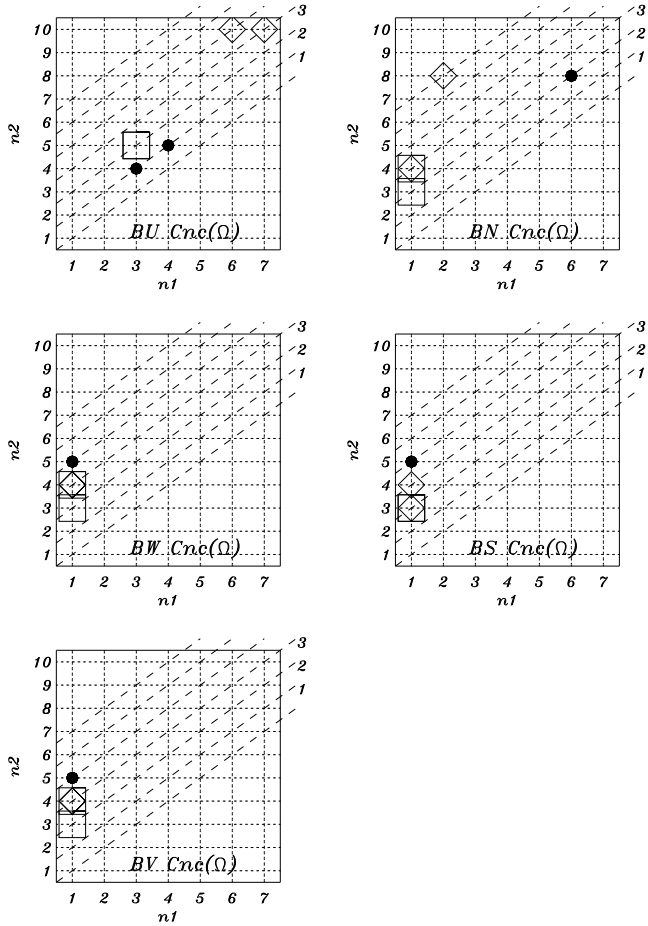}
  \caption{Same as Fig.~\ref{fig:GammesPrae}, but for models with effective temperatures 
           obtained from photospheric parameters that are corrected for rotational 
           effects (more details in text).}
   \label{fig:GammesPrae-rot}
\end{figure}
%

\subsection{Discussion of the present results}\label{ssec:Discussion}

Focusing on the present results, we come to the comparison between
observed and predicted ranges which is illustrated in Fig.~\ref{fig:GammesPrae}b.
Generally, the observed ranges are in good agreement with the theoretical predictions 
using $\anl=1.50$, whereas with $\anl=1.89$ the predicted ranges are in grave disagreement 
with the observations for the intermediate mass stars \BW\ and \BS.
In \michel\ this agreement (and disagreement) was observed for the set of four stars. 
However, in \michel\ the comparison between observations and theory was performed 
for a representative set of model solutions which were obtained independently 
for each individual star. On one hand, \michel\ reported that the observed and 
theoretical results could not be made to agree with $\anl=1.89$. On 
the other hand, their comparison was in reasonable agreement for models computed
with $\anl=1.50$ which, however, does not necessarily mean that they 
found one consistent model solution for all four objects (stars) all of which 
having the same age and metallicity. The results reported here, particularly 
the fact that we also do not find agreement between observations and theory for
models obtained with $\anl=1.89$, support the conclusions reported
by \michel, but limited to only one consistent model solution for all considered 
stars, which is the main objective of the present work. 

For the massive objects, \BU\ and \BN, the predicted unstable ranges are 
compatible with the observed results ($\pm1\,n$), for both $\anl=1.50$ and
$\anl=1.89$. The results for both objects are thus not sensitive to the value of $\anl$.
This is to be expected because these more massive stars have shallower 
outer convection zones and thus their structures are less sensitive to the assumed value 
of the mixing-length parameter.
In contrast, unstable ranges predicted for \BW\ \& \BS\ show a strong 
dependence on this parameter. The discrimination between the two $\anl$ values
reported in \michel\ is then partially reproduced. 
Finally, for the less massive object, \BV, the values for $\anl$ cannot be distinguished.
Nevertheless, the observed ranges agree with the theoretical
predictions within $\pm1\,n$. Therefore, in general, these results constitute a consistent solution
in terms of physics and cluster membership, and the observed and theoretical ranges of radial orders 
are in reasonable agreement for all the stars considered in this work. 

At this point, we should keep in mind that, as in \michel, 
the ranges of unstable modes
were estimated from non-rotating models for the observed stars. 
Therefore, we may wonder how accurately the so obtained theoretical ranges of 
unstable modes represent the observations of the rotating stars. 
However, the lack of theories describing the effect of rotation 
on mode stability makes it difficult to overcome this problem.
Nevertheless, assuming that mode stability depends predominantly
on the effective
temperature of the models \citep{Alosha75}, it is possible to make a rough 
estimate of the effect of varying the model's effective
temperature on the location of the range of unstable radial modes. 
The effective temperature varies if the photospheric 
 parameters are corrected for the effect of rotation 
(see Section~\ref{ssec:stelparam}). Keeping the remaining model parameters 
(mass, luminosity and chemical composition) constant we recalculate for each 
model the linear growth rates. The so obtained ranges of unstable radial 
modes are compared with the observations in Fig.~\ref{fig:GammesPrae-rot}.

In general, the results are similar to those in 
Fig.~\ref{fig:GammesPrae}b, but the ranges of unstable modes
are shifted. For $\anl=1.50$ the unstable ranges are shifted
to lower values in the case of \BU\ and \BN, but remain
almost unaltered for the remaining stars. On the other hand, 
for $\anl=1.89$ the unstable ranges are shifted to higher
values of $n$ for the more massive objects, and to lower values
for the rest of stars. 
In particular, the discrimination pattern of \BW\ and \BS\ in 
Fig.~\ref{fig:GammesPrae}b (obtained from the non-rotating equilibrium model 
parameters) is similar to the pattern of \BU\ and \BN\ in Fig.~\ref{fig:GammesPrae-rot} 
(obtained from model parameters for which $\teff$ is corrected for rotational
effects), which are more massive 
and hotter stars (see Table~\ref{tab:corrStd}).
The effective temperatures of \BW\ and \BS\ in Fig.~\ref{fig:GammesPrae-rot}
are similar to the effective temperature of \BV\ in Fig.~\ref{fig:GammesPrae}
and, as expected, the stability results are comparable.

\section{Conclusion}\label{sec:conclu}

In this work we studied ranges of unstable modes predicted by a stability analysis and compared the 
results with observations for a selected sample of \dss\ in the Praesepe cluster. With improved 
descriptions for rotational effects in {\bf both the evolutionary and  
adiabatic oscillations computations}, we searched for a single 
consistent solution in terms of physics and cluster membership in order to match reasonably
the observed and theoretically predicted ranges of overstable low-degree modes 
for all considered stars.
 
Consistent solutions were found for models with an age of 650 Myr, and which 
were computed with a mixing-length parameter of $\amlt=1.614$ and with a value
for $d_{\mathrm{ov}}$ that is compatible with the grid of solutions provided 
by \michel. In addition, for stellar models with $\log\,\teff=3.87$\,--\,$3.88$ 
a value of $\anl\simeq1.50$ leads to a reasonable agreement between 
theoretical predictions and observations of ranges of unstable modes, 
indicating that for these stars a smaller mixing-length parameter $\anl$ 
is required than suggested from a calibrated solar model. 
The need of a smaller value for $\anl$
than that from a calibrated solar model was also reported by
\citet{Pagoda05} for the $\delta$ Scuti star FG Vir.

Existing stability studies (including the present work) rely on calculations {\bf of
non-adiabatic quantities} without the inclusion of rotational effects; considering these 
effects in future investigations would lead to a substantial improvement of the stability 
analysis. In a first approximation we noticed an effect on mode stability through the 
change in effective temperature if it is corrected for rotational effects 
(see Section~\ref{ssec:stelparam}). The results presented here 
(Section~\ref{ssec:Discussion}) suggest, at least for some objects, a significant shift 
in the predicted ranges of unstable modes. Conclusions going beyond these findings, however, 
would require the development of a theory that describes the interaction between rotation 
and pulsation dynamics.

\section*{Acknowledgements}
First of all, we are grateful for the comments of the referee.
Then, JCS acknowledges support by the Instituto de 
Astrof\'{\i}sica de Andaluc\'{\i}a by an I3P contract
financed by the European Social Fund and from the Spanish 
Plan Nacional del Espacio under project ESP2004-03855-C03-01.
GH acknowledges support by the Particle Physics and Astronomy
Research Council of the UK.


\bsp

\label{lastpage}

\end{document}